# Raman Spectroscopy of Lithographically Patterned Graphene Nanoribbons

Sunmin Ryu,[1*] Janina Maultzsch,[2] Melinda Y. Han,[3] Philip Kim,[4] and Louis E. Brus[5]

[1]Department of Applied Chemistry, Kyung Hee University, Yongin, Gyeonggi 446-701, Korea

[2]Institut für Festkörperphysik, Technische Universität Berlin, 10623 Berlin, Germany

[3]Department of Applied Physics and Applied Mathematics, Columbia University, New York, NY 10027, USA. Current affiliation: National Renewable Energy Laboratory, Golden, CO, 80401, USA

[4]Department of Physics, Columbia University, New York, NY 10027, USA

[5]Department of Chemistry, Columbia University, New York, NY 10027, USA

*E-mail: sunryu@khu.ac.kr

**Abstract**

Nanometer-scale graphene objects are attracting much research interest because of newly emerging properties originating from quantum confinement effects. We present Raman spectroscopy studies of graphene nanoribbons (GNRs) which are known to have nonzero electronic bandgap. GNRs of width ranging from 15 nm to 100 nm have been prepared by e-beam lithographic patterning of mechanically exfoliated graphene followed by oxygen plasma etching. Raman spectra of narrow GNRs can be characterized by upshifted G band and prominent disorder-related D band originating from scattering at ribbon edges. The D-to-G band intensity ratio generally increases with decreasing ribbon width. However, its decrease for width < 25 nm, partly attributed to amorphization at the edges, provides a valuable experimental estimate on D mode relaxation length of <5 nm. The upshift in the G band of the narrowest GNRs can be attributed to confinement effect or chemical doping by functional groups on the GNR edges. Notably, GNRs are much more susceptible to photothermal effects resulting in reversible hole doping caused by atmospheric oxygen than bulk graphene sheets. Finally we show that the 2D band is still a reliable marker in determining the number of layers of GNRs despite its significant broadening for very narrow GNRs.

Graphene has attracted much interest as a novel two dimensional material with great potential in future applications such as flexible and transparent electrodes,[1-3] electrical devices,[4] ultrathin membranes,[5] and various nanocomposites[6] due to its remarkable material properties. In particular, nanometer-sized graphene objects (NGOs) are becoming more highlighted in further manipulating the inherent properties of bulk graphene sheets. Graphene nanoribbons (GNRs) have early been predicted to behave as a semiconductor with a bandgap which is determined by ribbon width and chirality.[7] The dimension-bandgap correlation has been experimentally confirmed for lithographically patterned GNRs.[8] Field effect transistors (FETs) made of chemically derived GNRs of width < 10 nm showed on-off current ratios of $10^7$ at room temperature.[9] Under appropriate in-plane electric fields, zigzag-edged GNRs have been proposed to behave as a half-metal which will be useful in spintronics on the nanometer scale.[10] As a networked nanostructure, graphene nanomeshes prepared by block copolymer lithography proved to carry 100 times larger current than individual GNR FETs[11] and also showed significant bandgaps.[12] Colloidal graphene quantum dots of variable size[13] and armchair-edged GNRs[14] were synthesized from molecular precursors and presented unique electronic and vibrational spectra, respectively.

Despite rising interest in quantum confinement effects of NGOs, however, optical characterization of their newly emerging properties has been rare.[13-16] In particular, Raman spectroscopy, widely used in characterizing bulk graphene sheets for their thickness,[17-19] structural integrity,[17, 20-23] and charge density,[24, 25] has not been systematically applied to NGOs, presumably due to insufficient Raman sensitivity.[8] As more NGOs are being studied, however, it has become ever important to have reliable tools to analyze their properties. It is not clear whether the 2D band[17-19] can still be used in differentiating single (1L) from double (2L) or multi-layered (nL) NGOs. Because of increased fraction of edge carbons in NGOs, the disorder-related D band intensity may not be useful in accessing their degree of structural defects. Functional groups at edges may further affect NGOs' electronic structures,[15, 26] in particular, the Fermi level by inducing extra charges,[27] which can be readily monitored by the positions of the G and 2D bands.[24, 25] Since theory predicts that degeneracy between transverse and longitudinal G bands of GNRs is to be lifted due to confinement,[28] it is of immediate importance to verify the size-effect on their lattice vibrations.[28-31]

Here we present systematic Raman spectroscopy studies of lithographically patterned GNRs. The nominal width ($w_{nom}$) ranges from 15 nm to 100 nm. While the relative intensity of the D band generally increases with decreasing width, it starts to decrease for the narrowest GNRs due to increasing fraction of disordered areas at edges. We also observe an upshift in the G band and drastic broadening in the G and 2D bands for the narrowest GNRs, which can be attributed to the size- and edge-effects. GNRs were also found to be much more sensitive to photothermally induced hole doping than bulk graphene sheets, which stresses that additional caution is required in diagnostic use of Raman spectroscopy for NGOs.

**RESULTS AND DISCUSSION**

The GNRs were prepared by lithographic patterning of hydrogen silsesquioxane (HSQ), a negative-tone resist, into protective etch masks followed by $O_2$ plasma etching of unprotected graphene area (see Methods).[8] Figure 1 shows AFM height images of sample **I** and sample **II** taken before (Fig. 1a) and after (Fig. 1b, 1c & 1d) $O_2$ plasma etching, respectively. The nominal widths of the GNRs were 15, 25, 50, and 100 nm for sample **I** and 15, 50, and 100 nm for sample **II**. Figure 2 presents the Raman spectra of GNRs as prepared (sample **I**). The spectra show the G band derived from the zone-center optical $E_{2g}$ phonon in graphene and two defect-induced bands denoted D and D$'$. Although the peak positions and intensities vary slightly from spot to spot, the overall spectral features were well reproduced within an array. The G band energy is virtually the same between the bulk patch and GNRs of width down to 25 nm. For the narrowest GNRs (15 nm), however, the G band upshifts by ~5 cm$^{-1}$ and broadens significantly. In addition, the G band was found to upshift further during repeated measurements, which suggests photo-induced effects that will be discussed later.

The two defect-induced bands at ~1330 cm$^{-1}$ (D band) and at ~1620 cm$^{-1}$ (D$'$ band) dominate the spectra as the ribbon width decreases. Figure 3 shows the integrated intensity ratio of the D band to G band ($I_D/I_G$) as a function of the ribbon width. As width decreases, the $I_D/I_G$ increases significantly for both 1L and 2L GNRs. The increasing D band intensity is due to the increasing fraction of edge carbons, which serve as defects by breaking the translational symmetry of the lattice.[17, 23] Since the employed e-beam lithography process is known to hydrogenate the basal plane forming C-H defects,[21] the $I_D/I_G$ ratio was also measured following two cycles of thermal dehydrogenation at 100$^o$C and 200$^o$C.[32] The annealing decreases the $I_D/I_G$ of the bulk graphene patch from 3.4 to 0.52, however, it barely affects that of narrow GNRs (width ≤ 50 nm), indicating that their basal plane C-H defects contribute much less than the ribbon edges in activating the D band (see Fig. S1 & S2, for Raman spectra of dehydrogenated GNRs). The 2L GNRs also show a similar increase in the $I_D/I_G$ ratio with decreasing ribbon width. Note that the patterning-induced basal plane hydrogenation is negligible in 2L graphene as was shown by zero $I_D/I_G$ ratio for the bulk.[21]

For polycrystalline graphite, the $I_D/I_G$ ratio correlates with effective domain size: with decreasing crystallite size ($L_a$), the ratio increases as $L_a^{-1}$ for $L_a$ > 2nm [33, 34] and decreases as $L_a^2$ below

~2 nm (solid lines in Fig.3).[23, 35] The decrease for $L_a$ < 2 nm was attributed to increasing fraction of amorphous sp$^2$ carbons which contribute to the G band but not the D band.[23, 35] We find that GNRs also show a similar scaling behavior: with decreasing width, the $I_D/I_G$ of 1L GNRs increases reaching a maximum at $w_{nom}$ of 25 nm and then decreases for the narrower GNRs ($w_{nom}$ = 15 nm). While 2L GNRs obey a similar rule, their $I_D/I_G$ ratios are smaller than those of 1L ones. If Wang *et al.*'s Raman measurement for ~2 nm-wide GNR[36] is included, it is clear that 2L GNRs also have a maximum $I_D/I_G$ ratio at width ≤ 15 nm.

The observed change in the scaling laws at a width of 15~25 nm provides an experimental confirmation of length scale for Raman scattering process of the D band.[37, 38] In the double resonance process for the D-mode phonon, an electron-hole pair generated by Raman excitation laser will propagate a certain distance, $\lambda$, during their lifetime which is limited by the uncertainty principle.[37, 39] Since the electron needs to be scattered off a defect to satisfy the momentum conservation rule,[40] the origin of the electron-hole pair needs to be within $\lambda$ from defects for the D band to be observed. The actual width ($w_{GNR}$) of GNRs prepared by identical methods is typically ~10 nm smaller than the nominal width due to over-etching underneath the HSQ etch mask.[8] The narrowest GNRs of $w_{nom}$ = 15 nm employed in the current study are likely ~5 nm wide. The extent of over-etching, however, should depend on various processing parameters. In addition, narrow areas (<2 nm, dark area of $w_{dis}$ in width in Fig. 3b) of the GNR edges are thought to be highly disordered due to O$_2$ plasma etching.[8, 41] While the G band can be observed from such strongly disordered sp$^2$ carbon networks, the D band cannot, due to the lack of six-fold rings.[23, 35] For a model GNR schematically shown in Fig. 3b, the central area (shown in red) located beyond $w_{dis} + \lambda$ from either edge does not contribute to the D band while still giving the G band signal. This reasoning leads to a conclusion that a maximum $I_D/I_G$ ratio will be obtained for a GNR of $w_{GNR} = 2\lambda + 2w_{dis}$. Given that $w_{nom}$ = 15~25 nm for maximum $I_D/I_G$ ratio in Fig. 3a and $w_{nom} = w_{GNR} + 10$ nm, we find that $\lambda$ is in the range of 1 ~ 5 nm, which is consistent with ~4 nm deduced from an argument based on lifetime and velocity of electrons.[37] Our result is also in good agreement with the Raman relaxation length (2 nm) determined for ion-bombarded bulk graphene sheets by Raman spectroscopy combined with STM.[38]

It is also notable that 1L GNRs give $I_D/I_G$ ratios much larger than 2L ones of the same width ($w_{nom}$) and graphite crystallites of $L_a = w_{nom}$: more appropriate graphene analogue for graphite crystallites of domain size $L_a$ will be graphene nanodisks of diameter $L_a$. However, the nanodisks will give higher $I_D/I_G$ than GNRs of width $L_a$ because of the former's larger fraction of edge carbons. This implies that NGOs obey the above empirical relationships[23, 33] qualitatively, but detailed relation is thickness-dependent. It is also to be noted that $I_D/I_G$ ratios of GNRs should depend on the polarization direction of the incoming and scattered light due to the 1-dimensional nature of the GNRs[42] and polarization-sensitivity of D band scattering off graphene edge.[37]

Raman scattering has been one of the most efficient and non-destructive tools to determine thickness of graphene samples.[17-19] Thus, it is an interesting question whether or not the optical diagnosis is still valid for NGOs. Fig. 4a shows 2D band Raman spectra of 1L GNRs as well as 1L bulk graphene. The 2D band of the GNRs can be well described with a single Lorentzian function and GNRs of $w_{nom} \geq 50$ nm are not distinguishable from the bulk graphene in terms of the 2D band's linewidth (30 ± 1 cm$^{-1}$). As can be seen in Fig. 5a, however, the 2D bands of sub-25 nm GNRs are noticeably broader than those of the rest. For 2L GNRs shown in Fig. 4b, the four sub-bands[17] of the 2D band persist even in the narrowest GNRs. However, the narrowest GNRs can be distinguished from the rest: (i) their peak frequency difference between the two central sub-bands is 4 cm$^{-1}$ larger than that of the rest (21±1 cm$^{-1}$); (ii) their sub-bands, except for 2D$_3$, are 5~10 cm$^{-1}$ broader than those of the wider GNRs. The broadening of the 2D bands (Fig. 5) and the G band (Fig. 2 & Fig. 6b) can be attributed to relaxation of momentum conservation rule in finite-size domains[23] and heterogeneity near GNR edges as will be discussed below. Despite the broadening, however, it is clear that Raman spectroscopy can be utilized in determining the number of layers in GNRs and possibly other NGOs as small as ~5 nm in lateral dimension.

Besides the broadening, the G band of the narrowest GNRs in Fig. 2 has ~5 cm$^{-1}$ higher frequency than wider ones (also see solid squares in Fig. 6a). Interestingly, we found that the unusually high G band frequency of the narrowest GNRs is partly due to oxygen in the air and can be partially reversed in an inert atmosphere: the G band downshifts by ~3 cm$^{-1}$ when measured in Ar gas (open diamonds in Fig. 6a). This indicates that the stiffening of the G band is largely caused by reversible binding of oxygen on GNRs. A similar sensitivity towards oxygen has been observed for thermally annealed graphene and pristine graphene with high G band frequency.[43] However, we also note that the G band frequency of the narrowest GNRs is still a few cm$^{-1}$ higher than wider GNRs even in inert Ar atmosphere, which may be attributed to quantum confinement effect in GNRs.[28-31] Theory predicts that new boundary conditions in GNRs lift the degeneracy of the $E_{2g}$ G mode.[28] As decreasing the width of GNRs, the transverse G mode subject to confinement across the GNRs blueshifts several times more than the longitudinal mode. The frequency splitting between the two orthogonal modes, pseudo-linearly decreasing with increasing width, amounts to 10~30 cm$^{-1}$ for GNRs of 2.5 nm.[31] Since the narrowest GNRs in the current study are thought to have a width of ~5 nm, non-negligible confinement effect can be expected. More specifically, unusually high G frequency and large linewidth of G and 2D bands found for the narrowest GNRs can be attributed to the size-effects.

Alternatively, functional groups at ribbon edges may be responsible for the anomalies in the narrow GNRs. As ribbon width decreases, edges represent a larger fraction of total carbons and thus become more important. In particular, our ribbons were patterned using oxygen plasma, a harsh oxidizing condition, which is known to form various oxygen-containing functional groups on graphite.[44] Since oxygen is more electronegative than carbon, such functional groups are expected to withdraw π electrons of GNRs (*i.e.,* dope with holes). Partially oxidized graphene sheets indeed show a significant upshift of the G-mode frequency.[20] Given the observed ~2 cm$^{-1}$ difference between the narrowest GNRs and the rest is due to edge functional groups, an average charge density difference of ~$2 \times 10^{12}$ cm$^{-2}$ can be estimated based on the electrical gating measurements.[24, 25] While we do not know the exact chemical identities and density of functional groups on the edges, we can estimate how much charge is transferred by one functional group with simple geometric considerations. Since an ideal armchair (zigzag) edge has 4700 (4000) edge carbons/μm, edge carbon density of 15 nm-wide GNRs is $6.2 \times 10^{13}$ ($5.3 \times 10^{13}$) cm$^{-2}$. For this model GNRs, ~0.03 charges would be required per edge carbon attached to one oxygen-containing functional group, which is not unrealistic.[45, 46] However, edges of lithographically patterned nanoribbons are very rough and contain highly disordered regions.[8] Moreover, charge density induced by edge functional groups should vary from edges to center of GNRs. The unusually broad G band of the narrowest GNRs can be attributed to such geometric imperfection and charge inhomogeneity.

In the presence of oxygen, further stiffening of the G band is induced by prolonged irradiation of the Raman excitation laser as can be seen in the progressive stiffening of the 25 nm-wide GNRs (Fig. 7a). The photo-induced stiffening in Ar gas, however, is several times less, which may be attributed to impurity-level residual oxygen in the Ar gas or other unknown origins. The narrowing of the G band (Fig. 7b) concurrent with the stiffening indicates that the changes can be explained by the charge doping theory:[24, 25] as the Fermi level is displaced from its neutrality point by charge doping, nonadiabatic electron-phonon coupling causes the G band to stiffen. Simultaneously, the linewidth of the G band decreases since Landau damping of the G phonon is not allowed when the Fermi level is located more than half of the G band energy away from the neutrality point.

The photo-induced changes in the G band are similar to what observed in thermally annealed bulk graphene sheets.[20, 21, 43, 47-50] To see the effect of thermal annealing on GNRs, the sample in Fig. 6 was re-investigated following annealing at 100°C in the air. It is notable that the G band frequency increased by 5~8 cm$^{-1}$ for the GNRs, while it remained within 1 cm$^{-1}$ for the bulk patch (blue circles in Fig. 6a). At the same time, the G band linewidth of the ribbons except the narrowest one decreased by 4~6 cm$^{-1}$ (Fig. 6b), which confirmed that the change is mostly driven by charge doping process. Stiffening of the D band (Fig. S3), thus 2D band, concurrent with that of the G band indicates that doped charges are mainly holes.[51] However, there was no significant change in the linewidth of the D band (Fig. S3).

The gas-sensitivity of the G band and D band can be seen respectively in Fig. 8 and Fig. S4, where the Raman spectra of 25 nm-wide GNRs were obtained in various gas environments before and after 100°C annealing. We found that the stiffening and narrowing of the G band and stiffening of D band correlate specifically with oxygen but not with other gases. The $O_2$-induced upshifts of 6 cm$^{-1}$ and 2 cm$^{-1}$ respectively for the G and D bands are reversible (also see Fig. S5). We also found that oxygen stiffens the 2D band as well as the G and D bands (Fig. S6). There was no noticeable difference between measurements in pure oxygen and air. Based on the identical sensitivity towards oxygen leading to the stiffening and narrowing of the G band, we conclude that the photo-induced effects are mainly caused by photo-thermal annealing of the GNRs.

It is interesting to note that thermally induced blueshifts in the G band frequency are significantly larger for GNRs than bulk graphene sheets (Fig. 6). Ryu et al. showed that the annealing-induced hole doping in bulk graphene sheets is directly caused by oxygen ($O_2$) and that the sensitivity to oxygen is greatly enhanced by thermally induced formation of nanometer-sized ripples.[43] Annealing can make graphene conform better to atomically rough $SiO_2$ substrate by driving out extraneous impurities trapped in the gap, which can lead to enhanced corrugation in graphene.[43] In an alternative explanation, heating causes graphene with negative thermal expansion coefficient (TEC) to shrink and slide against expanding $SiO_2$ substrate with positive TEC,[52] and the opposite differential expansion during cooling cycles can cause graphene to buckle in the out-of-plane direction forming ripples.[43] In GNRs, the thermal rippling can be readily explained by either of the two models: In the former, molecular species trapped underneath the GNRs can diffuse out very efficiently at elevated temperature due to their small width. In the latter, the total adhesive force between the GNRs and substrate, which resists sliding during heating, will be proportionally small for narrow GNRs.

The photo-induced effect is also dependent on the thickness of GNRs. Fig. 9 shows Raman spectra taken in the air for 1L and 2L 50 nm-wide GNRs before and after prolonged photoirradiation of the 514 nm Raman excitation laser. Upon the irradiation, the G band of 1L GNRs upshifted by 7 cm$^{-1}$, while that of 2L GNRs remained unchanged. This directly shows that the photothermal effect is much larger in 1L than 2L GNRs. Since thicker graphene sheets are stiffer,[53] formation of ripple requiring graphene sheets to deform in the out-of-plane direction should be less favored with increasing thickness. Similar thickness dependences of chemical activity have been observed in thermal oxidation,[20] hydrogenation,[21] photochemical[22] and electron transfer reactions[54] of bulk graphene sheets. We also note that the photoirradiation in Fig. 9 caused no significant structural defects. In fact, the D band intensity decreased following the irradiation indicating that the observed changes in the G band are not associated with formation of defects. The decrease in the D band intensity is due to the aforementioned photothermal removal of hydrogen atom defects bound to the basal plane of the GNRs.[21] Dehydrogenation itself was found to affect the G band frequency negligibly.[21]

**CONCLUSIONS**

In summary, we presented Raman spectroscopy studies for lithographically patterned GNRs of widths ranging from 15 to 100 nm. Despite significant broadening for narrow GNRs, the 2D band can still be used in determining the number of layers even for $w_{GNR}$ = ~5 nm ($w_{nom}$ = 15 nm). The Raman spectra of 1L GNRs are characterized by an upshift of the G band frequency and by an increase of $I_D/I_G$ ratio with decreasing width. The decrease in the $I_D/I_G$ ratio for the narrowest GNRs is attributed to amorphization near ribbon edges. The narrowest GNRs as prepared show significant broadening in both G and 2D bands and non-negligible blueshifts in the G band, which can be attributed to confinement effect and/or chemical doping from oxygen-containing groups at the GNR edges. Reversible chemical doping is considerably enhanced by laser irradiation in the presence of oxygen, which is attributed to structural deformation caused by photothermal annealing. 2L GNRs, however, show immunity towards photo-induced effects. Spectroscopic features of GNRs revealed in the current studies should contribute to characterization and application of various graphene nanostructures.

**METHODS**

The GNRs were prepared by lithographic patterning followed by oxidative etching.[8] 1L and 2L graphene sheets were deposited onto Si substrates with 300 nm-thick $SiO_2$ layer by mechanical exfoliation of Kish graphite. 40 nm-thick hydrogen silsesquioxane (HSQ), a negative-tone resist, was spin-coated onto the substrate. Following e-beam lithographic patterning of GNRs and subsequent development of the etch mask pattern, unmasked graphene area was etched by $O_2$ plasma. Since the HSQ overlayers are difficult to remove[8] and do not interfere with the optical measurements,[21] all the measurements were conducted with the GNRS capped by the HSQ layers. In sample **I** made of 1L graphene (Fig. 1), the nominal GNR widths in the four arrays under study were 15, 25, 50, and 100 nm; their length was 8~10 μm. To increase Raman signal, 6~10 GNRs of the same width were patterned in each array spanning 2 μm. Neighboring arrays were separated by 2 μm. All GNRs were oriented along the same direction. One large graphene patch (~8 μm across) served as a reference bulk graphene. Sample **II**, with 1L and 2L GNRs, was prepared similarly (Fig. 1). Raman measurements were performed with a home-built micro-Raman setup in backscattering geometry, using a He-Ne laser and an Ar ion laser for 633 nm and 514 nm excitations, respectively. The apparent diameters of focal spots were ~1.5 μm for 633 nm and ~1 μm for 514 nm. The spectral resolution limited by the overall instrumental response was 3.7 cm$^{-1}$ for 633 nm and 2.2 cm$^{-1}$ for 514 nm. While the polarization of the incoming light was parallel to the long axis of the GNRs, the scattered light was guided to a detector without further selection of polarization. All the measurements were done in ambient conditions except experiments where samples were placed in an environment-controlled optical cell. The pressure inside the cell was slightly above the atmospheric pressure and the typical gas flow rate was 0.20 L/min.

**Acknowledgements:** This research was supported by Basic Science Research Program through the National Research Foundation of Korea (NRF) funded by the Ministry of Education, Science and Technology (2010-0015363 and 2010-0028075) (to S.R.) J.M. thanks the Alexander-von-Humboldt foundation and the DFG under grant no. MA4079/7-1. This work was also supported at Columbia University by the National Science Foundation through the NSEC Program (CHE-06-41523) (to L.E.B.). PK and MYH acknowledge support from the EFRC Center for Re-Defining Photovoltaic Efficiency through Molecule Scale Control (award DE-SC0001085).

**Supporting Information Available**

Raman spectra of GNRs annealed at 100$^{o}$C and 200$^{o}$C, analysis of D band of GNRs of varying width, gas-sensitivity of D band of 25 nm-wide GNRs, analysis of G band of 25 nm-wide GNRs in response to various gases, O$_2$-sensitivity of 2D Raman bands of 25 nm-wide GNRs. This material is available free of charge *via* the Internet at http://pubs.acs.org.

**Figures**

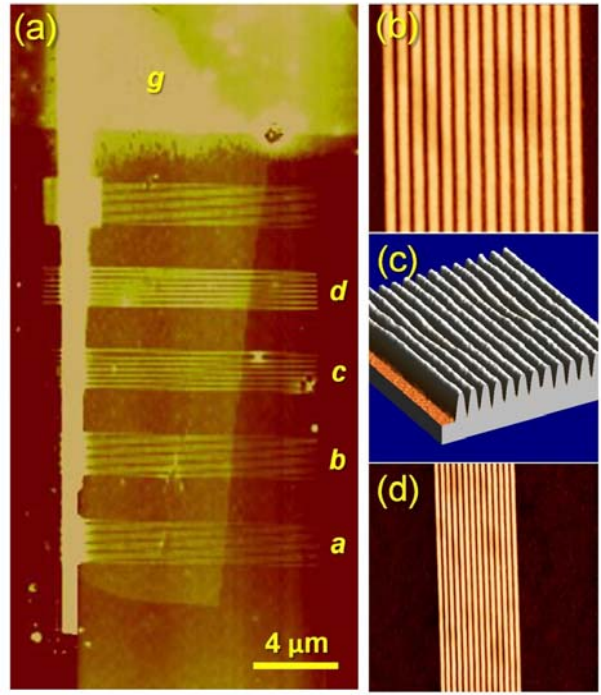

**Fig. 1.** (a) AFM height image of sample **I** with GNR sets of varying width (*a*: 15 nm, *b*: 25 nm, *c*: 50 nm, *d*: 100 nm, *g*: bulk graphene). The image was taken after patterning 40 nm-thick ribbon masks of HSQ and before removing unmasked graphene by oxygen plasma. Instrumental interference during AFM imaging caused Moire patterns along the GNRs, which blurred the observed topography. (b) AFM height image (2.5x2.5 μm$^2$) of 50 nm-wide GNRs in sample **II**. (c) 3-dimensional representation of the image in (b). (d) AFM height image (6x6 μm$^2$) of 50 nm-wide GNRs in sample **II**.

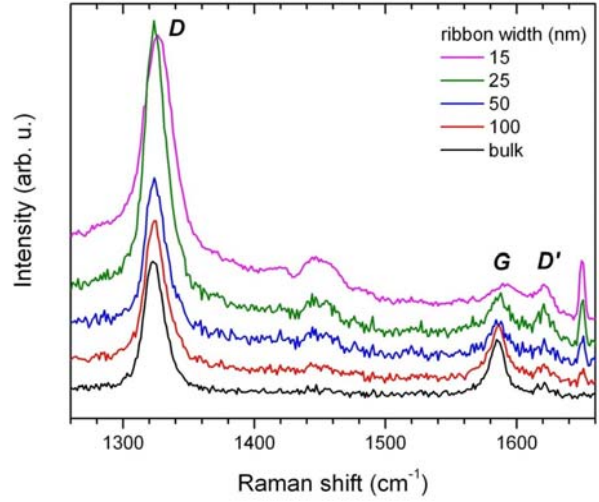

**Fig. 2.** Raman spectra of GNR sets *a, b, c, d, g* (see Fig. 1) taken for sample **I** as prepared without further treatment. Excited with laser wavelength λ = 632.8 nm. The spectra were offset for clarity. The bands at 1450 and 1650 cm$^{-1}$ are due to underlying Si and a plasma line of the excitation laser, respectively.

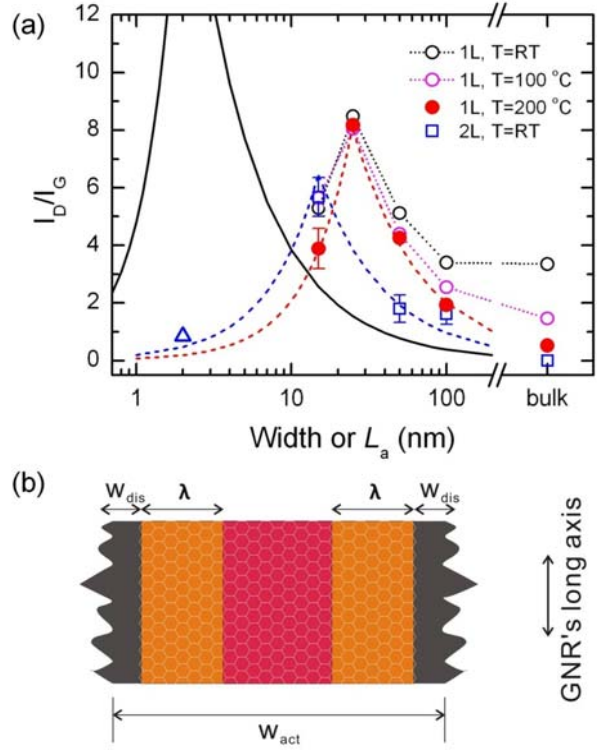

**Fig. 3.** (a) Integrated intensity ratio ($I_D/I_G$) of the D band to the G band as a function of the ribbon width ($w_{nom}$): 1L (circle, sample **I**) and 2L (square, sample **II**) GNRs; black circles were obtained from GNRs as prepared, magenta circles from GNRs annealed at 100°C, red solid circles from GNRs annealed at 200°C; blue squares from GNRs as prepared; open triangle is for 2 nm-wide 2L GNR from Ref. 36. Solid lines represent the empirical relations: $I_D/I_G \sim L_a^{-1}$ for $L_a > 2$ nm (Refs. 33 & 34) and $I_D/I_G \sim L_a^2$ for $L_a < 2$ nm (Ref. 23). They were calculated based on Refs. 34 & 23 using continuity at $L_a = 2$ nm (Ref. 23). Dashed and dotted lines are to guide eyes. (b) Schematic model of lithographically patterned GNR of $w_{act} = w_{nom} - 10$ nm. $\lambda$ and $w_{dis}$ are the relaxation length of the D band and width of disordered regions at edges, respectively.

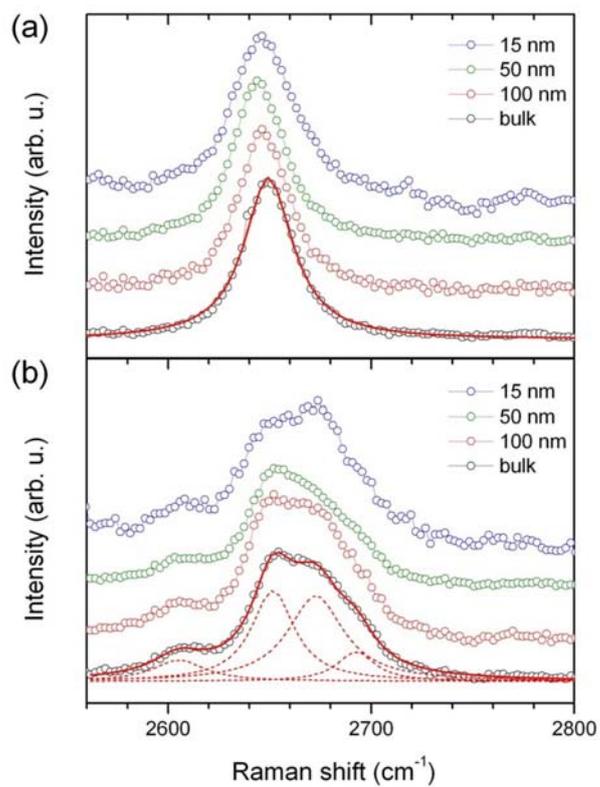

**Fig. 4.** 2D band Raman spectra of GNRs and bulk graphene sheets annealed in Ar at 100$^{o}$C for 1 hr: (a) 1L and (b) 2L (sample **II**). Red solid and dotted lines for bulk graphene sheets represent Lorentzian fits. ($\lambda_{excitation}$ = 632.8 nm).

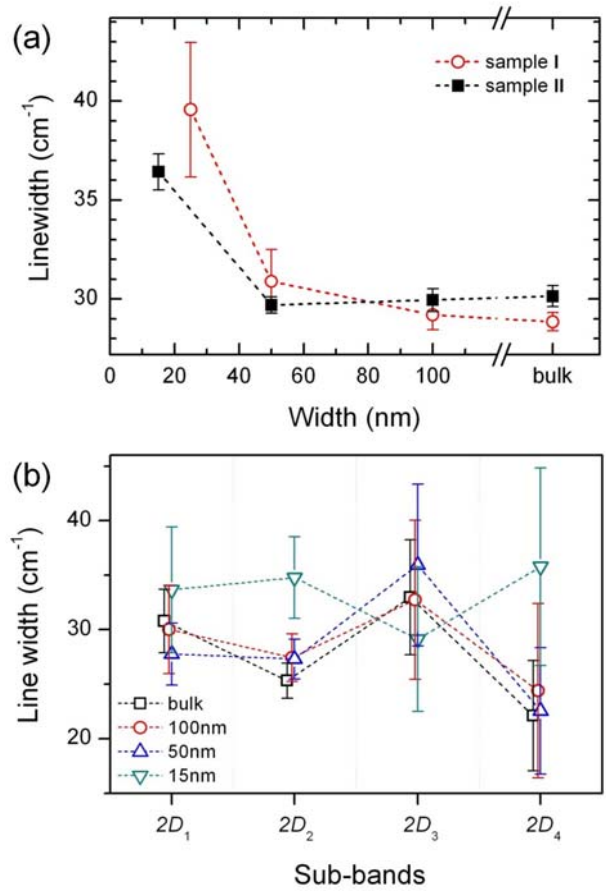

**Fig. 5.** (a) The linewidth of the 2D band of 1L GNRs as a function of ribbon width ($w_{nom}$). (b) The linewidth of the 2D band of 2L GNRs decomposed into four sub-bands.

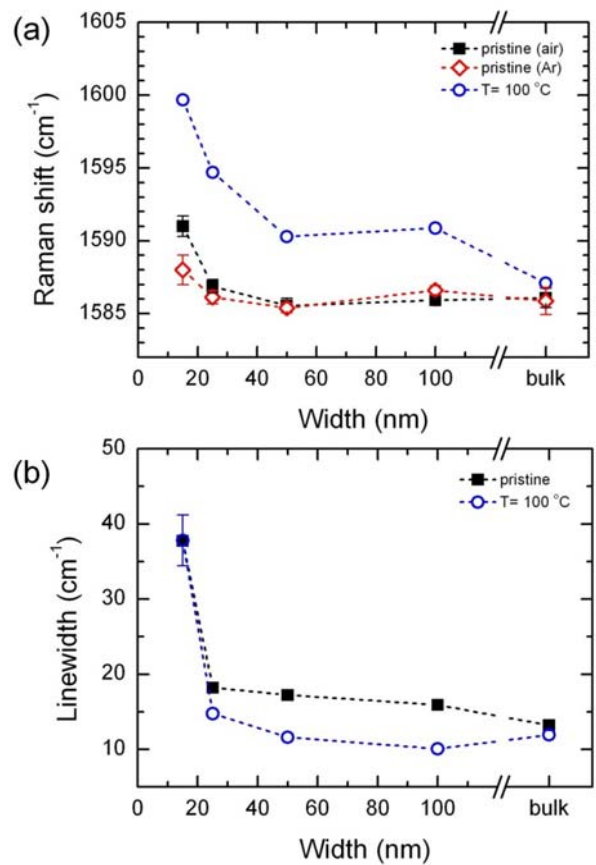

**Fig. 6.** The G band energy (a) and linewidth (b) of 1L GNRs as a function of ribbon width: (squares) as prepared, (circles) after annealing at 100°C. Diamonds in (a) were taken for the as-prepared sample under an Ar atmosphere instead of in air. The data were obtained from Fig. 2 and Fig. S1.

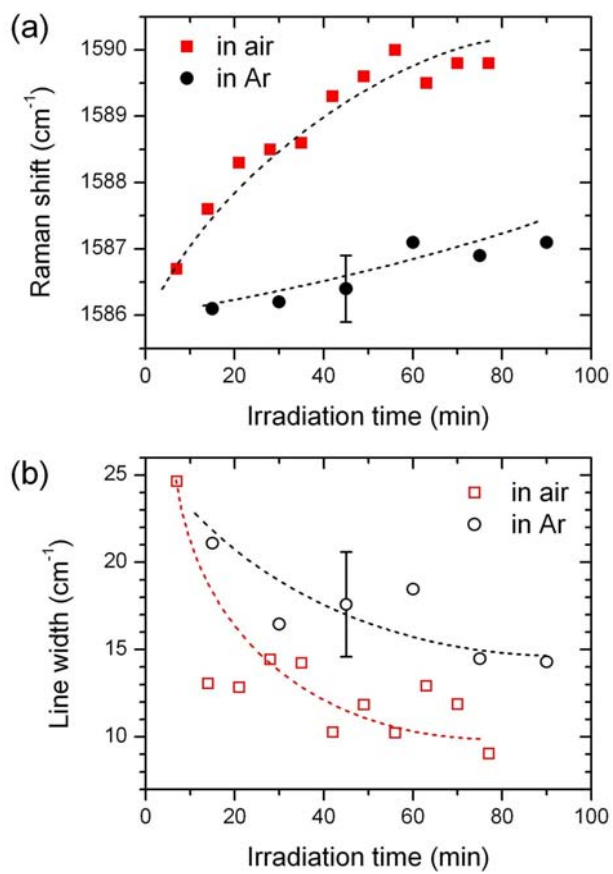

**Fig. 7.** The changes in the G band (a) energy and (b) linewidth of 1L 25 nm-wide ribbons (sample **I**, as prepared) caused by photoirradiation at 633 nm: (squares) in air, (circles) under an Ar atmosphere. The power density on the irradiated area (diameter ~1.5 µm) was 300 kW/cm$^2$. Dotted line is to guide eyes.

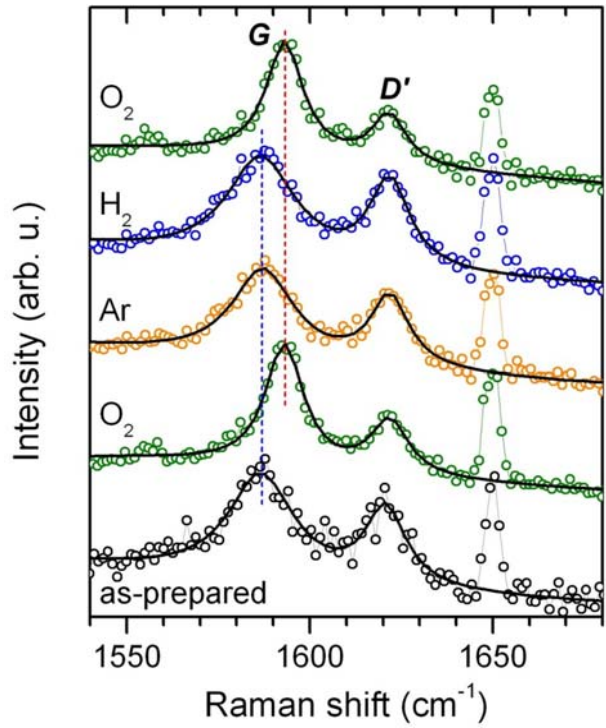

**Fig. 8.** Raman spectra of 25 nm-wide ribbons (sample **I**) obtained in various ambient gases. The bottom spectrum was taken in $O_2$ for as-prepared ribbons, and the rest four spectra were taken after annealing at 100°C for 2 hrs. The spectra are offset for clarity. Circles are experimental data and solid lines represent double Lorentzian fits with linear backgrounds. The plasma lines at ~1650 cm$^{-1}$ originating from the excitation laser were excluded from the fitting.

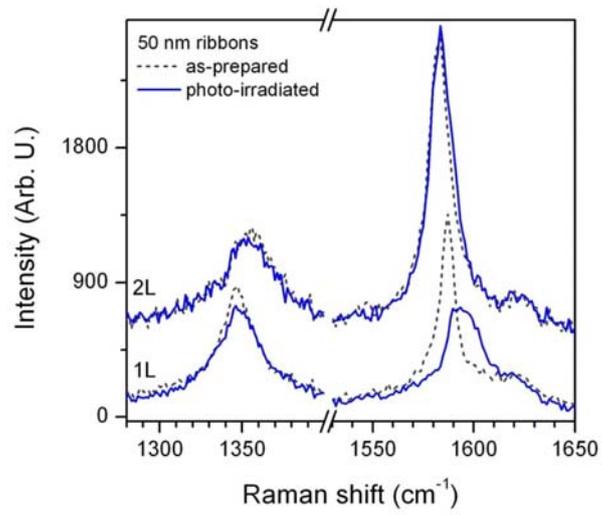

**Fig. 9.** The G band Raman spectra of 1L and 2L 50 nm-wide ribbons (sample **II**, as prepared) taken (dotted) before and (solid) after 10-min photo-irradiation at 514 nm. The power density on the irradiated area (diameter ~1 μm) was 300 kW/cm$^2$.

**TOC Figure**

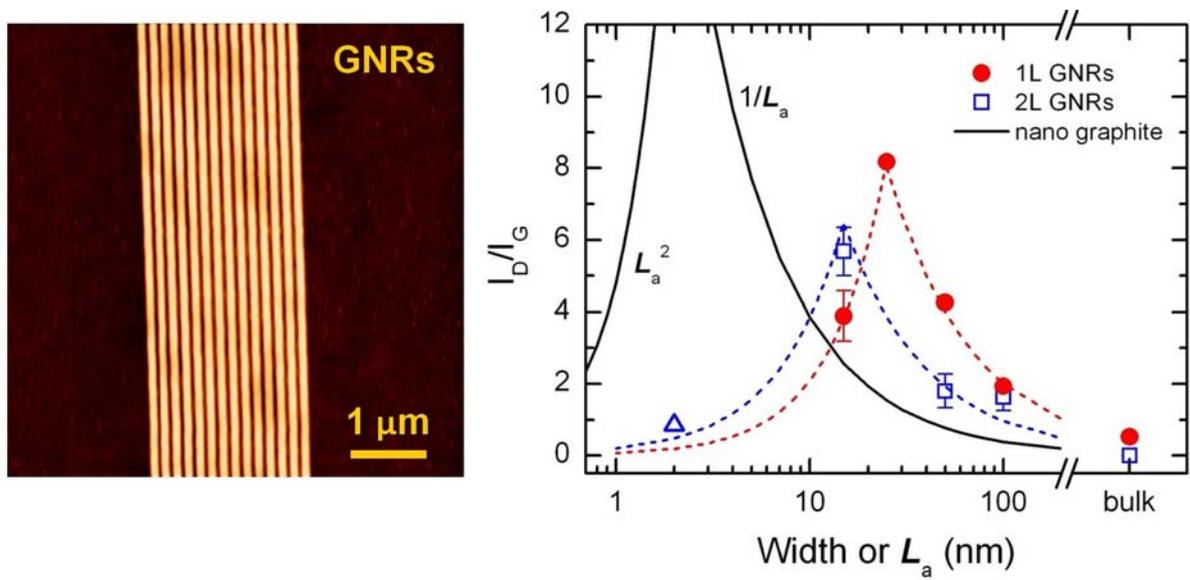